\documentclass[referee]{raa}            

\usepackage{graphicx,times}             
\usepackage{natbib}
\usepackage{amssymb,amsmath}
\bibpunct{(}{)}{;}{a}{}{,}
\usepackage[utf8x]{inputenc}
\usepackage[]{hyperref}
\usepackage[papersize={8.5in,11in}]{geometry}
\hypersetup{colorlinks = true, linkcolor = green, anchorcolor = red, citecolor = blue, filecolor = red, pagecolor = red, urlcolor = red}

\begin{document}

   \title{An investigation of planetary nebulae accompanying PG1159 central stars, based on Gaia DR2 measurements
}

   \volnopage{Vol.0 (20xx) No.0, 000--000}      
   \setcounter{page}{1}          

   \author{Ali, A.
      \inst{1}
   \and Wedad R. Alharbi
      \inst{2}
   }

   \institute{Department of Astronomy, Faculty of Science, Cairo University, 12613 Giza, Egypt;\\ {\it afouad@sci.cu.edu.eg}\\
        \and
             Department of Physics, College of Science, University of Jeddah, 23890 Jeddah, Saudi Arabia;\\ {\it wralharbi@uj.edu.sa}\\
\vs\no
   {\small Received~~20xx month day; accepted~~20xx~~month day}}

\abstract{This article discusses the physical and kinematical characteristics of planetary nebulae accompanying PG1159 central stars.  The study is based on the parallax and proper motion measurements recently offered by Gaia space mission. Two approaches were used to investigate the kinematical properties of the sample. The results revealed that most of the studied nebulae arise from progenitor stars of mass range; $0.9-1.75$\,M$_{\odot}$. Furthermore, they tend to live within the Galactic thick-disk and moving with an average peculiar velocity of $61.7\pm19.2$\,km\,s$^{-1}$ at a mean vertical height of $469\pm79$ pc. The locations of the PG1159 stars on the H-R diagram indicate that they have an average final stellar mass and evolutionary age of $0.58\pm0.08$\,M$_{\odot}$ and $25.5\pm5.3 \rm{x}10^3$ yr, respectively. We found a good agreement between the mean evolutionary age of the PG1159 stars and the mean dynamical age of their companion planetary nebulae ($28.0\pm6.4 \rm{x}10^3$ yr).
}

   \authorrunning{A. Ali \& W. R. Alharbi }            
   \titlerunning{PNe accompanying PG1159 central stars}  

   \maketitle

%
%
\section{Introduction} \label{introduction}

Gaia\footnote{More details regard Gaia mission are available at http://sci.esa.int/gaia/} is a space mission launched and operated by the European Space Agency (ESA) to provide a precise 3D map of our home galaxy (the Milky Way). Data was released in two versions; the first is Gaia DR1 in 2016 and the second is Gaia DR2 in 2018. Gaia provided astrometric and photometric data for about 1.7 billion sources, such as proper motion ($\mu$), parallax ($\varpi$) and magnitude in three photometric filters; G, G$_{BP}$, G$_{RP}$ \citep{Brown18}. The parallax measurements from Gaia DR2 and the forthcoming Gaia data release (Gaia\,DR3) represent a substantial step in the way of solving the chronic problem of finding reliable distances for the Galactic planetary nebulae (PNe).

Determining most, if not all, nebular parameters rely on their distances. It is known that only few PNe have trusted distances are determined by one of these individual methods; trigonometric parallax, spectroscopic parallax, cluster membership and angular expansion.  There are other less reliable statistical methods used to determine PNe distances, such as the relationships between the nebular ionized mass and both of its optical depth and its radius. For more discussion on these methods, see \citet{Ali15} and \citet{Frew16} and references therein. Although trigonometric parallax is one of the most reliable and direct methods for measuring PNe distances, it has been applied to only few PNe (e.g. \citealt{Harris07, Acker98, Pottasch&Acker98}) before Gaia era. Distances for most stars in Gaia DR2 could not be calculated directly from their parallax angles since these angles were measured using a complicated iterative technique that includes different assumptions. \citet{Bailer-Jones18} applied a correct inference procedure to represent the nonlinearity of the transformation and asymmetry of the resulting probability distribution. The uncertainty in the distance assessment is represented by the lower and upper limits of asymmetric confidence interval. Therefore, in this study, we adopt the distances taken from \citet{Bailer-Jones18}.

\begin{table}
\begin{center}
\caption[]{The basic data of the PNe-PG1159 sample.}\label{Table1}
\scalebox {0.49}{
\begin{tabular}{lllccccccccccccccclcccc}

\hline
\\
\multicolumn{13}{c}{Planetary nebulae parameters} && \multicolumn{6}{c}{Central stars parameters}\\
\\
\cline{1-13} \cline{15-20}\\
PN designation	&$L(^{\circ})$ & $B(^{\circ})$ &	\multicolumn{3}{c}{$\alpha$} &  &	\multicolumn{3}{c}{$\delta$}	&	$\theta(")$	&	$V_{los}$ (${\rm km\,s^{-1}}$)& $V_{exp}$ (km\,s$^{-1}$)	&&	Gaia DR2 ID	&	$D$ (pc)	&	$PM_{\alpha} (mas/yr)$	&	$PM_{\delta} ( mas/yr) $	&	$m_G$	&	$BP-RP$	\\
\cline{4-6} \cline{8-10}
\\
\hline \\
Sh\,2-68	&	30.673	&	6.279	&	18	&	24	&	58.43	&		&	0	&	51	&	36.02	&	200	&				&	5	$\pm$	0.5	&&	4276328581046447104	&	400	&	-19.06	$\pm$	0.12	&	-56.06	$\pm$	0.13	&	16.4	&	0.28	\\
A\,43	    &	36.062	&	17.62	&	17	&	53	&	32.28	&		&	10	&	37	&	24.21	&	40	&	-42	$\pm$	11.5	&	40	$\pm$	2	&&	4488953930631143168	&	2187	&	-5.61	$\pm$	0.11	&	1.03	$\pm$	0.1	&	14.7	&	-0.28	\\
NGC\,6852	&	42.59	&	-14.528	&	20	&	0	&	39.21	&		&	1	&	43	&	40.92	&	14	&	-11	$\pm$	5	&	43.1	$\pm$	4.3	&&	4237745794618477440	&	3099	&	-1.38	$\pm$	0.27	&	-4.75	$\pm$	0.2	&	17.9	&	-0.34	\\
Sh\,2-78	&	46.832	&	3.845	&	19	&	3	&	10.08	&		&	14	&	6	&	58.78	&	298	&	42	$\pm$	1.7	&	20	$\pm$	2	&&	4506484097383382272	&	629	&	-1.11	$\pm$	0.21	&	-4.21	$\pm$	0.22	&	17.6	&	-0.05	\\
A\,72	    &	59.795	&	-18.729	&	20	&	50	&	2.05	&		&	13	&	33	&	29.52	&	68	&	-59	$\pm$	23	&				&&	1761341417799128320	&	1395	&	1.46	$\pm$	0.11	&	-6.28	$\pm$	0.09	&	16	&	-0.54	\\
NGC\,6765	&	62.458	&	9.557	&	19	&	11	&	6.56	&		&	30	&	32	&	43.68	&	17	&	-64	$\pm$	14.6	&	35	$\pm$	3.5	&&	2039515046433996544	&	3476	&	-0.61	$\pm$	0.25	&	-4.3	$\pm$	0.34	&	17.5	&	-0.2	\\
NGC\,7094	&	66.778	&	-28.202	&	21	&	36	&	52.97	&		&	12	&	47	&	19.1	&	50	&	-101	$\pm$	30.8	&	38	$\pm$	2	&&	1770058865674512896	&	1548	&	3.61	$\pm$	0.09	&	-10.6	$\pm$	0.09	&	13.5	&	-0.45	\\
Kr\,61	    &	70.524	&	11.007	&	19	&	21	&	38.94	&		&	38	&	18	&	57.22	&	48	&	−25.4	$\pm$	2	&	67.6	$\pm$	6.8	&&	2052811676760671872	&	3306	&	-0.1	$\pm$	0.29	&	-3.78	$\pm$	0.26	&	18.3	&	-0.47	\\
MWP\,1	    &	80.356	&	-10.409	&	21	&	17	&	8.28	&		&	34	&	12	&	27.42	&	336	&				&	30	$\pm$	3	&&	1855295171732158080	&	495	&	-5.47	$\pm$	0.07	&	10.44	$\pm$	0.08	&	13	&	-0.59	\\
Jacoby\,1   	&	85.366	&	52.349	&	15	&	21	&	46.56	&		&	52	&	22	&	3.87	&	330	&				&	30	$\pm$	3	&&	1595941441250636672	&	734	&	-3.96	$\pm$	0.1	&	11.26	$\pm$	0.14	&	15.6	&	-0.66	\\
K\,1-16	    &	94.025	&	27.428	&	18	&	21	&	52.11	&		&	64	&	21	&	53.41	&	57	&				&	22.5	$\pm$	2.3	&&	2160562927224840576	&	1986	&	-3.27	$\pm$	0.13	&	-3.05	$\pm$	0.11	&	15	&	-0.63	\\
Jn\,1	    &	104.208	&	-29.642	&	23	&	35	&	53.32	&		&	30	&	28	&	6.34	&	163	&	-67	$\pm$	30	&	15	$\pm$	1.5	&&	2871119705335735552	&	808	&	2.97	$\pm$	0.13	&	0.84	$\pm$	0.09	&	16	&	-0.53	\\
NGC\,246    	    &	118.863	&	-74.709	&	0	&	47	&	3.34	&		&	-11	&	52	&	18.98	&	122	&	-46	$\pm$	3.6	&	39.5	$\pm$	4	&&	2376592910265354368	&	506	&	-16.96	$\pm$	0.21	&	-9	$\pm$	0.13	&	11.8	&	-0.65	\\
NGC\,650    	    &	130.934	&	-10.504	&	1	&	42	&	19.66	&		&	51	&	34	&	31.55	&	70	&	-19	$\pm$	1.2	&	39	$\pm$	3.9	&&	406328439057955968	&	2873	&	-0.25	$\pm$	0.81	&	-4.16	$\pm$	0.47	&	17.3	&	0.02	\\
IsWe\,1	    &	149.715	&	-3.398	&	3	&	49	&	5.91	&		&	50	&	0	&	14.9	&	363	&	-2	$\pm$	0.5	&	12	$\pm$	1.2	&&	250358801943821952	&	441	&	18.01	$\pm$	0.17	&	-7.43	$\pm$	0.13	&	16.5	&	-0.26	\\
JnEr\,1     	    &	164.806	&	31.181	&	7	&	57	&	51.62	&		&	53	&	25	&	16.94	&	190	&	-84	$\pm$	8.8	&	24	$\pm$	2.4	&&	936605992140011392	&	979	&	0.14	$\pm$	0.15	&	-0.72	$\pm$	0.12	&	17.1	&	-0.63	\\
A\,21	    &	205.139	&	14.241	&	7	&	29	&	2.71	&		&	13	&	14	&	48.59	&	316	&	29	$\pm$	5.2	&	32	$\pm$	3.2	&&	3163546505053645056	&	531	&	-2.72	$\pm$	0.14	&	-8.6	$\pm$	0.13	&	16	&	-0.59	\\
Lo\,3 	    &	258.068	&	-15.748	&	7	&	14	&	49.42	&		&	-46	&	57	&	39.21	&	47	&				&	15.6	$\pm$	1.6	&&	5509004952576699904	&	1807	&	0.84	$\pm$	0.2	&	4.46	$\pm$	0.23	&	16.8	&	-0.04	\\
Lo\,4	    &	274.309	&	9.112	&	10	&	5	&	45.79	&		&	-44	&	21	&	33.52	&	20	&	33	$\pm$	20	&				&&	5414927915911816704	&	3180	&	-6.15	$\pm$	0.12	&	0.3	$\pm$	0.13	&	16.6	&	-0.47	\\

\hline
\end{tabular}}
\end{center}
\end{table}

All born stars with initial masses $\leq 8.0\,M_{\odot}$ will end their lives as white dwarf (WD) stars. The evolution of the star between the asymptotic giant branch (AGB) and the WD phase takes a short time compared to the preceding evolutionary phases. During this short transition, all stars that start their evolution with hydrogen and helium burning shells (post-AGB) will end their lives with carbon-oxygen WD cores \citep{Werner&Herwig06}. The central stars (CSs) of PNe can be divided into two main groups according to the hydrogen abundance in their atmospheres \citep{Mendez91}. The first group “hydrogen-rich” has relative hydrogen abundance close to the cosmic value, while the second group “hydrogen-deficient” has high abundances of helium and carbon with a tiny (or free) amount of hydrogen. The stellar spectra of the latter group are dominated by broad and intense emission lines typical of Wolf-Rayet [WR] stars, frequently of [WC] subtype and sometimes of [WO]. The spectra of some hydrogen-deficient stars reveal the presence of helium, carbon and oxygen absorption lines. This class was named PG1159 after the detection of its prototype star PG 1159-035. In addition, there is a small set of stars that has the same properties of PG1159 class but their spectra show some hydrogen absorption lines are named hybrids-PG1159 \citep{Napiwotzki&Schoenberner91}. The PG1159 stellar class is characterized by effective temperature (T$_{\rm eff}$) ranging from 7500\,K to 25000\,K and logarithmic surface gravity (Log\,g) from $5.5\, \rm{cm}\,s^{-2}$ to $8.0\, \rm{cm}\,s^{-2}$ \citep{Lobling19}.  \citet{Werner&Herwig06} reported an evolutionary sequence for hydrogen-deficient C-rich stars as follows: $AGB \rightarrow [WC] \rightarrow PG 1159 \rightarrow DO$.

The main objective of the present work is to determine the kinematical and physical parameters of PNe accompanying PG1159 CSs. The rest of the article is structured as follows. Section \ref{sample} presents the data sample. Sections \ref{phy-char} and \ref{Kin-char} inspect the physical and kinematical properties of the selected sample, respectively. Section \ref{PG1159} discusses the status of the four nebulae A\,21, IeWe\,1, Sh\,2-78 and NGC\,650, while we drew our conclusions
in Section \ref{conclusions}.

\section{The sample} \label{sample}
An update of the CSs catalog of \citet{Weidmann11} has been recently published by \citet{Weidmann20}. The new catalog contains the spectral classification of 620 CSs of confirmed and probable PNe compared with 492 in the old catalog. Examining the new catalog and other literature, we obtained an initial sample of 22 PNe accompanying PG1159 central stars (PNe-PG1159). The CSs of A\,43, NGC\,7094 and Sh\,2-68 have spectral characteristics of hybrids-PG1159. The CSs of A\,30, A\,78 and NGC 2371-72 are classified as stars in the transition phase between [WC] and PG1159. The spectrum of [WC]-PG1159 class shows properties of both [WC] and PG1159 stars. These three nebulae were excluded from our sample, and hence our final sample consists of 19 PNe-PG1159. Table \ref{Table1} summarizes the basic data of the PNe-PG1159 sample. It shows the PN name, galactic ($L$, $B$) and equatorial ($\alpha$, $\delta$) coordinates, angular radius ($\theta$), line of sight velocity ($V_{los}$), expansion velocity ($V_{exp}$),  Gaia\,DR2 designation, distance ($D$), proper motion ($PM_{\alpha}$, $PM_{\delta}$), G-magnitude ($m_G$), and color index ($BP-RP$). The PNe coordinates and angular radii were collected from the SIMBAD database and Table 10 \citep{Frew16}, respectively, while the PNe radial and expansion velocities are from \citet{Acker92} and \citet{Frew16}. The CSs parameters are gathered from the Gaia DR2 catalog.
It should be noted that the central star is the source of the UV radiation that ionizes the gas within the nebular shell, and hence it appears as a blue star. The measurements of $BP-RP$ listed in Table 1 showed, that all PG1159 stars are blue stars except Sh\,2-68 is red. The reason behind the observed red color of Sh\,2-68 may be attributed to either the high reddening along its line of sight direction or to visible light being dominated by its main sequence close binary companion.

\begin{table}
\centering
\caption{The physical parameters of the sample, where $R$, $T_{dyn}$, $V_{LSR}$ and  $|V_{p}|$ are the PN radius, dynamical age, line of sight velocity corrected for LSR and absolute peculiar velocity, respectively. $T_{ev}$ and $M_f$ are the evolutionary age and final mass of the central star, respectively.} \label{Table2}
\scalebox{0.70}{
\begin{tabular}{lllcccc}
\hline
PN designation	&	$R$ (pc)			&	$T_{dyn} \rm{x} 10^3$ (yr)	&	$T_{ev} \rm{x} 10^3$ (yr) & $M_f$ (M$_{\odot})$	&	$V_{LSR}$ (km\,s$^{-1}$)			&	$|V_{p}|$ (km\,s$^{-1}$)			\\
\hline
Sh\,2-68	&	0.39	$\pm$	0.04	&		75.8	$\pm$	11.0	&	104.0	$\pm$	15.0	&	0.63	$\pm$	0.11	&			\\
A\,43	&	0.42	$\pm$	0.07	&			10.4	$\pm$	1.8 &	8.3	$\pm$	1.7	&	0.57	$\pm$	0.06	&	-27.08	$\pm$	7.41	&		90	$\pm$	11	\\
NGC\,6852	&	0.20	$\pm$	0.09	&		4.6	$\pm$	2.1	&				&	0.53	$\pm$	0.00	&	4.13	$\pm$	1.88	&	99	$\pm$	56	\\
Sh\,2-78	&	0.91	$\pm$	0.12	&		44.4	$\pm$	7.3	&	37.3	$\pm$	8.0	&	0.65	$\pm$	0.09	&	56.35	$\pm$	2.29	&		24	$\pm$	2.4	\\
A\,72	&	0.46	$\pm$	0.08	&								&	67.0	$\pm$	13.0	&				&	-43.38	$\pm$	17.03	&		87	$\pm$	17	\\
NGC\,6765	&	0.29	$\pm$	0.14	&		8.0	$\pm$	4.0	&				&	0.53	$\pm$	0.00	&	-49.35	$\pm$	11.21	&		134	$\pm$	36	\\
NGC\,7094	&	0.38	$\pm$	0.05	&		9.7	$\pm$	1.4   &	8.6	$\pm$	1.5
	&	0.53	$\pm$	0.00	&	-86.23$\pm$	26.27	&		125	$\pm$	26	\\
Kr\,61	&	0.77$\pm$0.34	&		11.1$\pm$5.0	&				&				&	-11.13	$\pm$	0.88	&		73	$\pm$	15	\\
MWP\,1	&	0.81$\pm$0.08	&		26.3$\pm$5.6	&	3.4	$\pm$	0.8	&	0.58	$\pm$	0.00					&				\\
Jacoby\,1	&	1.17$\pm$0.17	&		38.3$\pm$10.0	&	5.8	$\pm$	1.8	&	0.58	$\pm$	0.06			&				\\
K\,1-16	&	0.54$\pm$0.08	&		23.7$\pm$4.2	&	12.5	$\pm$	2.0	&	0.53	$\pm$	0.00				&				\\
Jn\,1	&	0.64$\pm$0.08	&		41.6$\pm$6.7	&	10.6	$\pm$	1.8	&   0.56	$\pm$	0.04	&	-56.68$\pm$	25.38	&		76$\pm$26	\\
NGC\,246	&	0.30$\pm$0.03	&		7.4	$\pm$1.1	&	8.3	$\pm$	2.1	&	0.59$\pm$0.04	&	-38.82$\pm$3.04	&		45	$\pm$	3.1	\\
NGC\,650	&	0.97$\pm$0.41	&		24.4$\pm$11.0	&	9.9	$\pm$	1.9	&	0.56$\pm$	0.04	&	-15.63$\pm$	0.98	&		3$\pm$	5	\\
IsWe\,1	&	0.77$\pm$0.08	&		63.1$\pm$22.0	&	53.4$\pm$2.6	&	0.65$\pm$0.17	&	-3.36$\pm$1.12	&		12$\pm$1.1	\\
JnEr\,1  	&	0.90$\pm$0.13	&	36.8$\pm$7.0	&	44.5	$\pm$	5.4	&	0.56	$\pm$	0.04	&	-86.16	$\pm$	8.99	&	87$\pm$	9	\\
A\,21	&	0.81$\pm$0.09	&		24.8	$\pm$	2.6	&	22.5	$\pm$	7.0	&	0.58	$\pm$	0.06	&	17.85$\pm$3.22	&		24$\pm$	3.2	\\
Lo\,3 	&	0.41$\pm$0.10	&		25.8	$\pm$	6.6	&	10.0	$\pm$	2.2	&	0.53	$\pm$	0.00				&				\\
Lo\,4	&	0.31$\pm$0.08	&								&	1.5	$\pm$	0.5	&	0.74$\pm$0.04	&	23.79$\pm$14.42	&		54$\pm$14	\\

\hline
\end{tabular}}
\end{table}

\section{Physical characteristics of the sample} \label{phy-char}
It is noticeable from Table \ref{Table1} that about half the sample objects are located at high galactic latitudes with a\ mean absolute value of $20.5^{\circ}$. Moreover, the determined mean vertical height ($Z$) is $469\pm79$\,pc (Table \ref{Table4}). The latter result indicates that most PNe-PG1159 reside inside the Galactic thick-disk, which has a mean vertical height of $510\pm40$\,pc \citep{Carollo10}, and hence they are frequently members of Galactic population II.
In Table \ref{Table2}, we present the PN radius which calculated from its angular radius and distance. The dynamical (kinematical) age of the PN was calculated from its derived radius and expansion velocity. The analysis of these two parameters showed that this class of PNe are of large sizes and long ages. Throughout the PN evolution, its size and mass increase while its density decreases.  The results reported in Table \ref{Table2} show an average radius of $0.60\pm0.13$\,pc, which is approximately six times the standard PN radius ($0.1$\,pc; \citealt{Pottasch83}) and an average dynamical age of $28.0\pm6.4\rm{x}10^3$\,yr, which is nearly three times the standard value ($10.0\rm{x}10^3$\,yr; \citealt{Pottasch83}). The expansion velocity of PN is not constant but it varies during the nebular evolution as a result of the variation of the stellar wind parameters throughout PN dynamical evolution. Therefore, we should take the dynamical age with caution throughout discussing the physical parameters of PNe.

\begin{figure*}
{ \begin{tabular}{@{}ccc@{}}
	\includegraphics[scale=0.45]{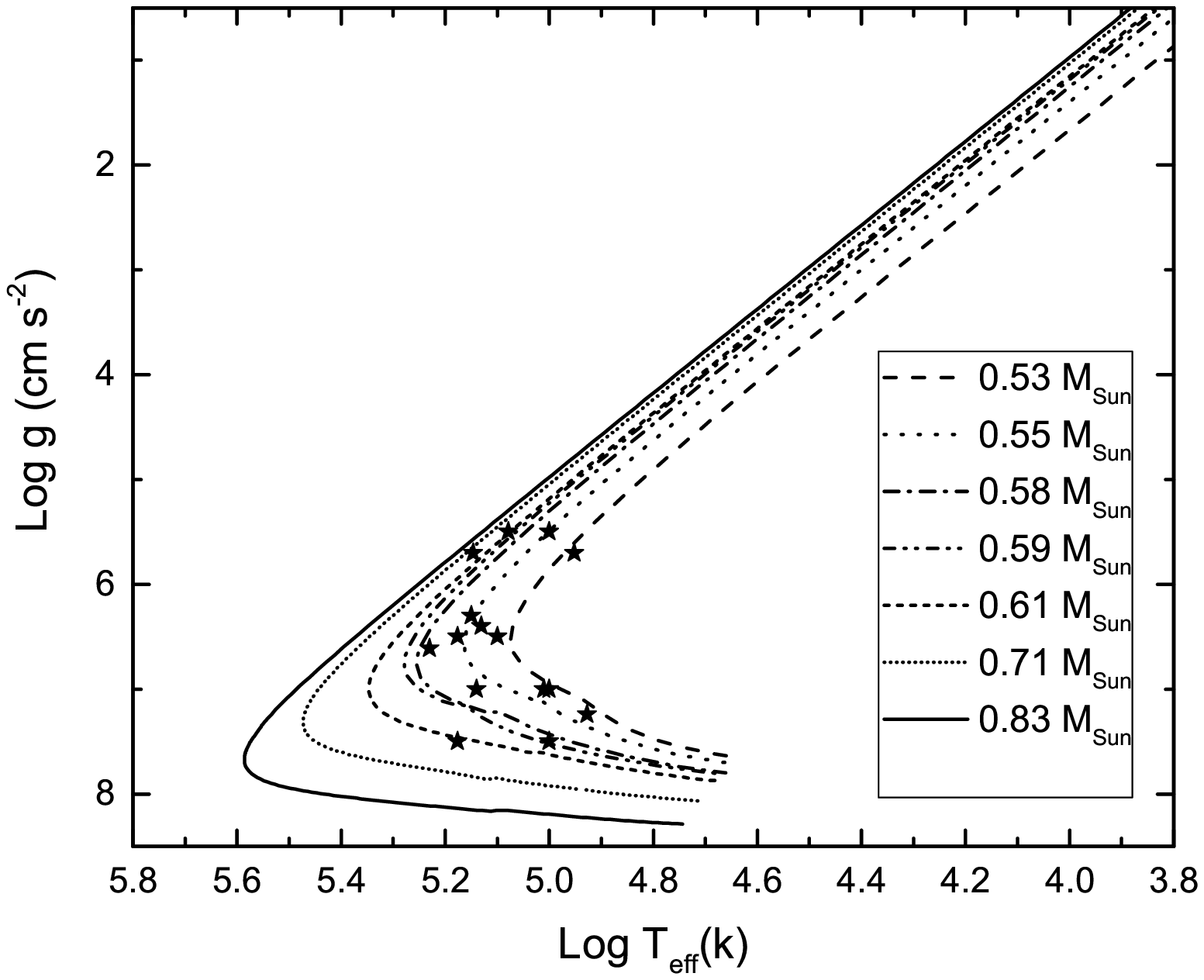} &
     \includegraphics[scale=0.45]{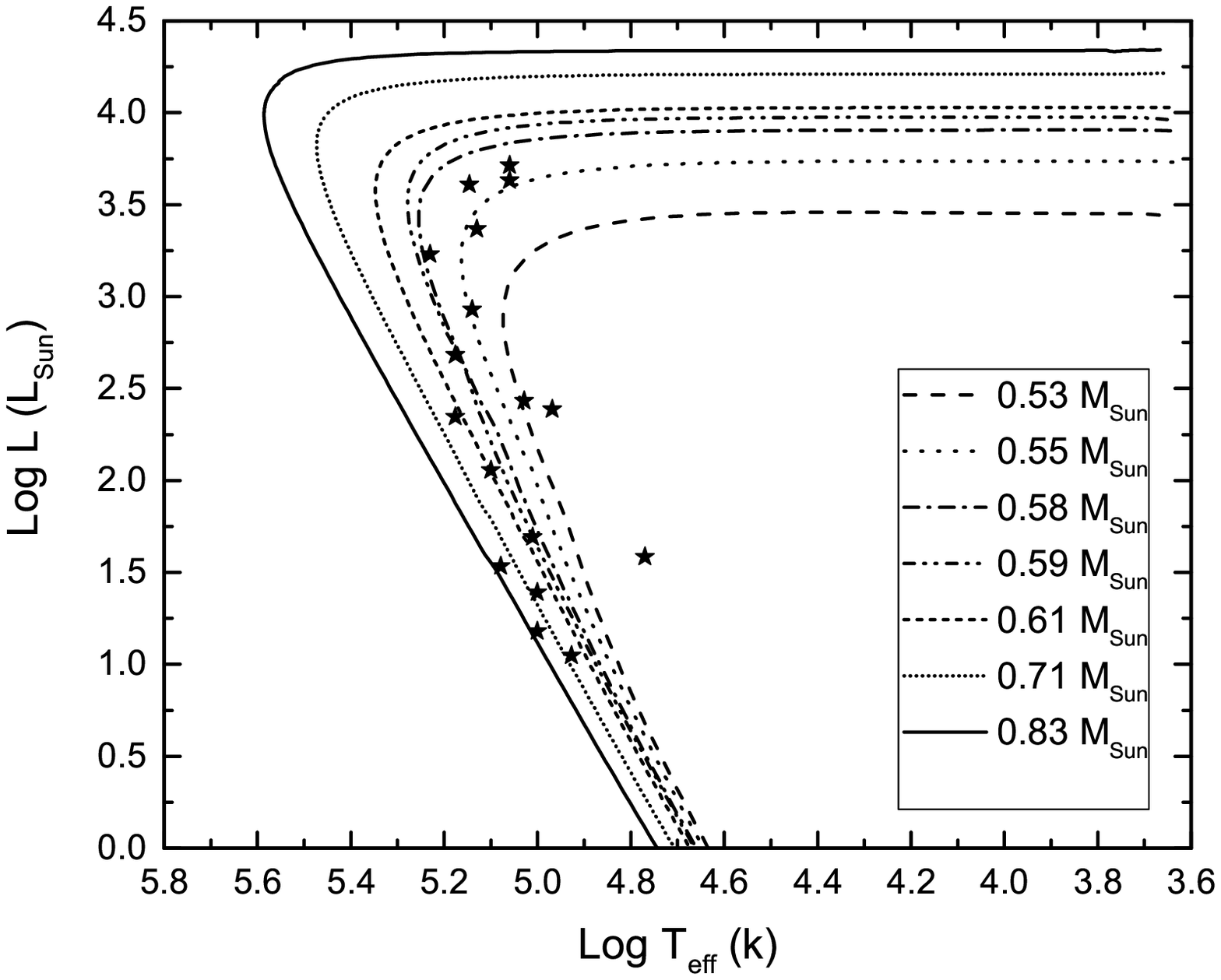}
		\end{tabular}}
\caption{Location of PG1159 central stars on the H-R diagram. The hydrogen-burning evolutionary tracks overlaid on the diagram are taken from \citep{Bertolami16} with initial metallicitiy of $Z=0.001$. The left and right panels show the position of PG1159 stars on Log\,T$_{\rm{eff}}$\,-\,Log\,g and Log\,T$_{\rm{eff}}$\,-\,Log\,L charts, respectively. The final mass of each model star was given on both charts. To clarify the figure, the  evolutionary age isochrones of various stellar models and PNe names were omitted} \label{Figure1}
\end{figure*}

\begin{table}
\centering
\caption{The effective temperature, surface gravity, and luminosity of the PG1159 central stars.} \label{Table3}
\scalebox{0.85}{
\begin{tabular}{lccc}
\hline
PN designation		& Log\,T$_{\rm{eff}}$ (K)	& Log\,g (cm\,s$^{-2}$)	&		Log\,L (L$_{\odot}$)	\\
\hline
Sh\,2-68	&	4.93	& 7.24	&		1.05	\\
A\,43	    &	5.00	& 5.50	&		3.29	\\
NGC 6852	&	4.77	&		&	1.59	\\
Sh\,2-78	&	5.00	& 7.50	&		1.39	\\
A\,72	    &	5.03	&		&	2.43	\\
NGC\,6765	&	4.97	&		&	2.39	\\
NGC\,7094	&	4.95	& 5.70	&		3.40	\\
Kr\,61	    &		    &		&		\\
MWP\,1	    &	5.23	& 6.61	&		3.23	\\
Jacoby\,1	&	5.18	& 7.50	&		2.34	\\
K\,1-16	    &	5.13	& 6.40	&		3.37	\\
Jn\,1	    &	5.18	& 6.50	&		2.68	\\
NGC\,246	&	5.15	& 5.70	&		3.61	\\
NGC\,650	&	5.14	& 7.00	&		2.93	\\
IsWe\,1	    &	5.00	& 7.00	&		1.18	\\
JnEr\,1  	&	5.01	& 7.00	&		1.69	\\
A\,21	    &	5.10	& 6.50	&		2.05	\\
Lo\,3 	    &	5.15	& 6.30	&			\\
Lo\,4	    &	5.08	& 5.50	&		1.53	\\
\hline
\end{tabular}}
\end{table}

Table \ref{Table3} lists the effective temperature, surface gravity and luminosity of the PG1159 central stars.  The effective temperature and surface gravity were extracted from \citet{Weidmann20} and \citet{Werner&Herwig06}, whereas the luminosity was derived from the Gaia distance and CS visual magnitude, corrected for interstellar extinction, following \citet{Frew08}. The PN extinction coefficients are gathered from \citet{Frew16}. Figure \ref{Figure1} shows the H-R diagram of the CSs sample in which the left and right panels illustrate the locations of the PG1159 stars, that have available data, on the Log\,T$_{\rm{eff}}$-Log\,g and Log\,T$_{\rm{eff}}$-Log\,L charts, respectively. The H-burning post-AGB tracks of diverse masses and initial metallicitiy Z=0.001 (\citealt{Bertolami16}) were overlaid on both charts . These tracks describe the theoretical evolution of seven model stars with different initial masses (0.9, 1.0, 1.25, 1.75, 2.0, 2.5, and 3\,M$_{\odot}$) from the onset of post-AGB phase until the end of white dwarf cooling sequence phase. The final stellar masses of the diverse stellar models (0.53, 0.55, 0.58, 0.59, 0.61, 0.71, and 0.83\,M$_{\odot}$) are illustrated in the figure. Most PG1159 stars reside close to the tracks with stellar final masses less than 0.59\,M$_{\odot}$.  The average final mass of the PG1159 stars is $0.58\pm0.08$\,M$_{\odot}$, which is slightly smaller than the value (0.62\,M$_{\odot}$) derived by \citet{Werner&Herwig06}. From the age isochrones, we inferred the evolutionary age(T$_{ev}$) of each star. The estimated final mass and evolutionary age of the PG1159 stars are given in Table \ref{Table2}. The results show acceptable agreement between the evolutionary age of each PG1159 star and its companion PN dynamical age (Table \ref{Table2}). The derived mean evolutionary age of the PG1159 stars, $25.5\pm5.3\rm{x}10^3$ yr, is slightly lower than the calculated mean nebular dynamical age ($28.0\pm6.4\rm{x}10^3$\,yr).

\section{Kinematical characteristics of the sample} \label{Kin-char}
To investigate the kinematical properties of PNe-PG1159, we used two approaches. The first is to build the $\sqrt{U^2_{LSR}+W^2_{LSR}}$ - $V_{LSR}$ diagram “Toomre diagram”, while the second is to calculate the peculiar velocity of the sample.
\begin{table}
\centering
\caption{The space coordinates and velocity components of the sample.} \label{Table4}
\scalebox{0.70}{
\begin{tabular}{lccccccc}
\hline
PN designation	&	$X$ (pc)	&	$Y$	(pc)&	$Z$	(pc)&	$U$	(km\,s$^{-1}$)		&	$V$	(km\,s$^{-1}$)		&	$W$	(km\,s$^{-1}$)		&	$V_s$	(km\,s$^{-1}$)		\\
\hline
Sh\,2-68	&	7258	&	203	&	44	&				&				&				&				\\
A\,43	&	5915	&	1227	&	662	&	-27.0	$\pm$	7.0	&	179.0	$\pm$	28.0	&	49.0	$\pm$	10.0	&	69.4	$\pm$	27.0	\\
NGC\,6852	&	5391	&	2030	&	777	&	47.2	$\pm$	25.3	&	163.4	$\pm$	63.2	&	-5.5	$\pm$	3.3	&	73.9	$\pm$	61.0	\\
Sh\,2-78	&	7171	&	458	&	42	&	48.0	$\pm$	1.9	&	247.0	$\pm$	16.9	&	7.0	$\pm$	50.0	&	55.5	$\pm$	72.0	\\
A\,72	&	6935	&	1142	&	448	&	1.3	$\pm$	1.7	&	154.5	$\pm$	41.5	&	-4.3	$\pm$	3.2	&	65.7	$\pm$	41.0	\\
NGC\,6765	&	6015	&	3039	&	577	&	43.0	$\pm$	41.0	&	141.0	$\pm$	32.0	&	-25.0	$\pm$	9.0	&	93.4	$\pm$	33.0	\\
NGC\,7094	&	7062	&	1254	&	731	&	-0.2	$\pm$	0.3	&	96.4	$\pm$	18.9	&	-7.6	$\pm$	8.1	&	123.8	$\pm$	19.0	\\
Kr\,61	&	6518	&	3059	&	631	&	54.0	$\pm$	27.9	&	188.6	$\pm$	32.0	&	-22.1	$\pm$	8.6	&	66.3	$\pm$	32.0	\\
MWP\,1	&	7518	&	480	&	90	&				&				&				&				\\
Jacoby\,1	&	7564	&	447	&	581	&				&				&				&				\\
K\,1-16	&	7724	&	1758	&	915	&				&				&				&				\\
Jn\,1	&	7772	&	681	&	400	&	13.0	$\pm$	29.0	&	166.0	$\pm$	70.0	&	40.0	$\pm$	18.0	&	68.4	$\pm$	68.0	\\
NGC\,246	&	7664	&	117	&	488	&	62.0	$\pm$	3.0	&	219.0	$\pm$	32.0	&	46.0	$\pm$	4.0	&	77.2	$\pm$	30.0	\\
NGC\,650	&	9451	&	2134	&	524	&	23.0	$\pm$	15.0	&	198.0	$\pm$	65.0	&	-45.0	$\pm$	20.0	&	55.1	$\pm$	63.0	\\
IsWe\,1	&	7980	&	222	&	26	&	-9.0	$\pm$	0.4	&	191.0	$\pm$	7.2	&	18.0	$\pm$	0.8	&	35.3	$\pm$	7.2	\\
JnEr\,1  	&	8408	&	220	&	507	&	79.0	$\pm$	8.0	&	203.0	$\pm$	19.0	&	-36.0	$\pm$	4.0	&	88.5	$\pm$	18.0	\\
A\,21	&	8066	&	-218	&	131	&	-11.6	$\pm$	2.5	&	197.0	$\pm$	15.7	&	-0.7	$\pm$	0.1	&	25.8	$\pm$	16.0	\\
Lo\,3 	&	7960	&	-1701	&	490	&				&				&				&				\\
Lo\,4	&	7364	&	-3131	&	504	&	-64.0	$\pm$	16.1	&	179.0	$\pm$	77.1	&	-38.0	$\pm$	10.8	&	85.0	$\pm$	72.0	\\
\hline
\end{tabular}}
\end{table}

\subsection{Toomre diagram} \label{Toomre-diagram}
Toomre diagram has been introduced by \citet{Bensby03} and \citet{Bensby10} to carefully investigate the Galactic population of F and G dwarfs. To construct this diagram, we derived the space velocity components ($U$, $V$, and $W$) and their uncertainties following the procedure suggested by \citet{Johnson&Soderblom87}. We ignored the errors associated with the equatorial and galactic coordinates, which have an insignificant effect on the final results. The uncertainties in the $U$, $V$, and $W$ components were estimated by propagating the errors in the distance, radial velocity, and proper motion. $U$ and $V$  denote  the Galactic center and Galactic rotation directions, respectively, whereas $W$ indicates the direction perpendicular to the Galactic disk. The velocity components were corrected for the local standard of rest (LSR) assuming that the solar peculiar motion ($U_{\odot}$,$V_{\odot}$, $W_{\odot}$) = (9.0, 11.0, 6.0)\,km\,s$^{-1}$. The space velocity component $V$ that presented in Table \ref{Table4}, is corrected further for the rotational solar velocity $V(R_{\odot})=220$\,km\,s$^{-1}$. To locate the sample objects on the Toomre diagram, we calculated their total space velocities $V_S$ ($\sqrt{U^2_{LSR}+V^2_{LSR}+W^2_{LSR}}$).

\begin{figure}
\centering
	\includegraphics[scale=0.7]{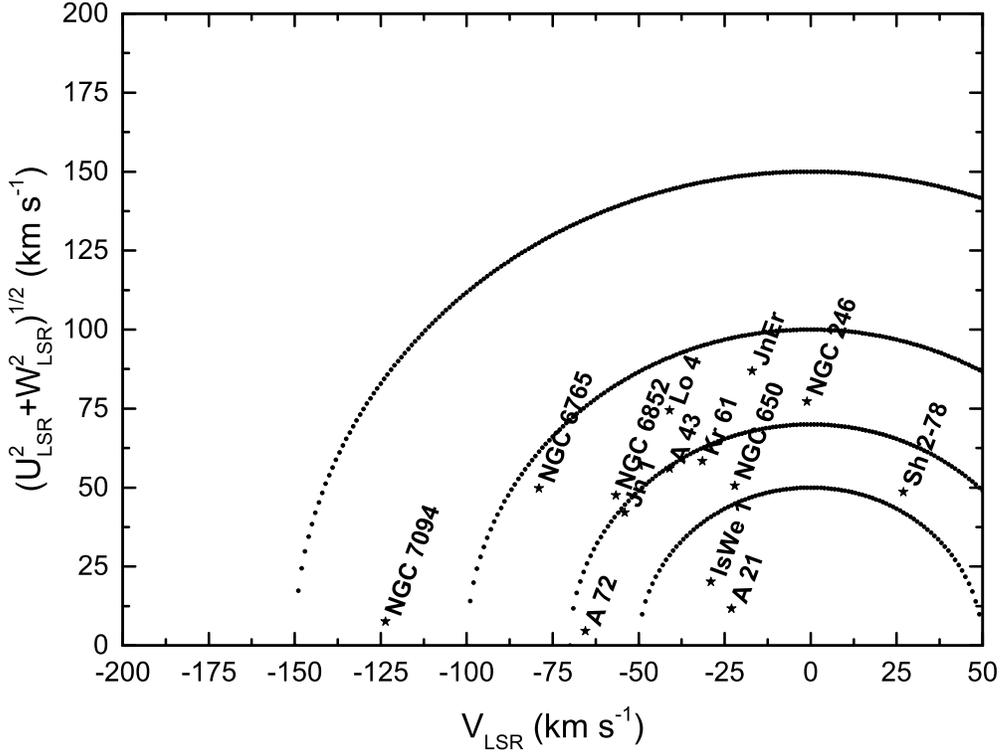}
	\caption{Toomre diagram of our sample. The space velocity components $U$, $V$, and $W$ are corrected for the LSR. Dotted semi-circular lines demonstrate constant total space velocity values:  50, 70, 100, and 150 km\,s$^{-1}$. The error bars for each PN were removed to clarify the figure.} \label{Figure2}
\end{figure}

In the Toomre diagram, stars with $V_S \leq 50\,{\rm km\,s^{-1}}$  locate in the Galactic thin-disk and those with $200 \geqslant V_S \geqslant 70$\,km\,s$^{-1}$ reside in the Galactic thick-disk, while stars with $V_S > 200\,{\rm km\,s^{-1}}$ settle in the Galactic halo. Objects with $70 \geqslant V_S \geqslant 50$\,km\,s$^{-1}$ have similar likelihood to exist either in the Galactic thin or thick-disk. Figure \ref{Figure2} displays that A\,21 and IsWe\,1 are associated with the Galactic thin-disk, while NGC\,246, NGC\,6765, NGC\,6852, JnEr\,1, NGC\,7094, A\,43 and Lo\,4 are associated with the Galactic thick-disk. The nebulae NGC\,650, Kr\,61, A\,72, Sh\,2-78 and Jn\,1 have equal probability to belong to either the Galactic thin or thick-disk.  In addition to the space velocity components of the studies objects, we presented, in Table \ref{Table4}, their space coordinates $X$ (towards the Galactic anti-centre), $Y$ and $Z$ (towards the Galactic North Pole). We adopted here, the solar space coordinates $X_{\odot}=7.6$ kpc, $Y_{\odot}=0.0$ and $Z_{\odot}=0.0$.

\subsection{Peculiar velocity} \label{Peculiar-velocity}
Peculiar velocity ($V_p$) expresses the difference between the line of sight velocity corrected for LSR and the velocity derived from the Galactic rotation curve, assuming the object has a circular orbit \citep{Quireza07}. \citet{Maciel&Dutra92} proposed the concept of peculiar velocity to distinguish between the different types of Peimbert classification \citep{Peimbert78}. According to this classification, \citet{Quireza07} consider all PNe moving with $|V_p| \geq 60$\, km\,s$^{-1}$ as high velocity nebulae of Peimbert type III while that moving with $|V_p| < 60$\,km\,s$^{-1}$ as low velocity nebulae of Peimbert types I and II. Although Peimbert types are mainly based on the chemical composition of the PNe, they show further variation in their kinematical properties. Types I and II PNe are members of Galactic population I that live in the Galactic thin-disk, while type III PNe are members of Galactic population II that live in the Galactic thick-disk. In general, PNe arise from a wide stellar mass domain ranging from 0.8\,M$_{\odot}$ to 8.0\,M$_{\odot}$. Peimbert type III nebulae mostly originate from the lower progenitor mass stars compared to types I and II. As a result of scarce information regarding the observed chemical abundances of nearly all PNe in the sample, it is difficult to get their Peimbert types. Based on the modest abundances of few chemical elements in some objects and using the Bayes theorem, \citet{Quireza07} were able to predict the Peimbert type III for NGC\,6765 and type IIa for Jn\,1, NGC\,246 and NGC\,650.

The absolute peculiar velocities are calculated following \citet{Quireza07} and the results are presented in Table \ref{Table2}. We found eight objects out of 14 possessing $|V_p| \geq 60$\, km\,s$^{-1}$ and two objects possessing $|V_p| \geq 45$\,km\,s$^{-1}$. This refers to a moderate tendency for the nebular sample to be of Peimbert type III, and hence, they mostly belong to the Galactic thick-disk and  originated from low mass stars. The derived peculiar velocities agree with the results deduced from the Toomre diagram (Section \ref{Toomre-diagram}) where the objects of small peculiar velocities, e.g., IsWe\,1 ($|V_p| = 12\pm1.1$\,km\,s$^{-1}$) and A\,21 ($|V_p| = 24\pm3.2$\,km\,s$^{-1}$) occupy the Galactic thin-disk, whereas objects of high velocities, e.g., NGC\,6765 ($|V_p| = 134\pm36$\,km\,s$^{-1}$) and NGC\,7094 ($|V_p| = 125\pm26$\,km\,s$^{-1}$) occupy the Galactic thick-disk.

It is noticeable that although  NGC\,650 has insignificant peculiar velocity, it has high galactic latitude and vertical height. Further, its measured total space velocity indicates that the location of this object is in the area of overlap between the thin and thick Galactic disk ($70 \geqslant V_S \geqslant 50$\,km\,s$^{-1}$). The former results promoted the likelihood membership of NGC\,650 to the Galactic thick-disk instead of the thin-disk. Taking into consideration the galactic heights and peculiar velocities, we found that Kr\,61, A\,72 and Jn\,1 (that also lie in the overlapping area between the Galactic thin and thick disk on the Toomre diagram) have much tendency to belong to the Galactic thick-disk, whereas Sh\,2-78 to the Galactic thin-disk.

\section{Are A\,21, I\lowercase{e}W\lowercase{e}\,1, s\lowercase{h\,2-78} and NGC\,650 really associate with PG1159 CS\lowercase{s}?} \label{PG1159}

The nebulae A\,21, IeWe\,1, Sh\,2-78 and NGC\,650 hold faint central stars of visual magnitudes  $17.94\pm0.03$ \citep{Pena97}, $16.56\pm0.10$ \citep{Ishida87}, $17.78\pm0.03$ \citep{Cappellaro90} and $17.70\pm0.20$ \citep{Napiwotzki93}, respectively. These four central stars are classified as PG1159 type by \citet{Napiwotzki92} and \citet{Napiwotzki93} using low-resolution spectra. \citet{Pena97} noticed shallow and wide hydrogen and helium absorption lines in the optical spectrum of A\,21 central star indicating a hot hydrogen-rich white dwarf of DAO spectral type, on reverse to the prior classification as hydrogen-deficient of PG1159 spectral type \citep{Napiwotzki93}. Unfortunately, no recent medium or high dispersion spectra are available for the other three stars to affirm their early PG1159 classification.

\section{Conclusions} \label{conclusions}
We have conducted an analysis for the available planetary nebulae that hosting central stars of spectral type PG1159. The kinematical and physical characteristics of PNe-PG1159 objects were discussed in detail. The leading results clearly point out that the majority of these nebulae are belonging to the Galactic thick-disk population. Furthermore, they are evolved planetary nebulae of Peimbert type III that originated from low mass progenitor stars. The mean dynamical age of the nebular sample declares good agreement with the mean evolutionary age of their PG1159 central stars, and both are about three times the standard value. A further argument confirming the aging of these nebulae is their mean large size, which is about six times the common nebular size. Finally, it is worth noting that the sample utilized in this analysis is statistically small to set up solid conclusions.

\begin{acknowledgements}
The authors would like to thank the anonymous referee for his valuable comments that enhanced the paper. Further, we appreciate the useful remarks of  Dr. Awad, Z. and Dr. Snaid, S., which helped us improve the reading of the paper. This work has made use of data from the European Space Agency (ESA) mission Gaia, processed by the Gaia Data Processing and Analysis Consortium(DPAC). This research has made use of the SIMBAD database, operated
at CDS, Strasbourg, France.
\end{acknowledgements}

\bibliographystyle{raa}
\bibliography{RAA-2020-0416}

\end{document}